\documentclass[twocolumn]{aastex631}
\usepackage{lineno}

\newcommand\latex{La\TeX}

\shorttitle{X-ray flaring of HD 54786}
\shortauthors{Bhattacharyya et al.}
\graphicspath{{./}{figures/}}

\begin{document}

\title{Decoding the X-ray flare from MAXI J0709-159 using optical spectroscopy and multi-epoch photometry}

\author[0000-0002-1920-6055]{Suman Bhattacharyya}
\affiliation{Department of Physics and Electronics, CHRIST (Deemed to be University), Hosur Main Road, Bangalore, India\\}

\author[0000-0002-7254-191X]{Blesson Mathew}
\affiliation{Department of Physics and Electronics, CHRIST (Deemed to be University), Hosur Main Road, Bangalore, India\\}

\author[0000-0003-1795-3281]{Savithri H Ezhikode}
\affiliation{Department of Physics and Electronics, CHRIST (Deemed to be University), Hosur Main Road, Bangalore, India\\}

\author{S Muneer}
\affiliation{Indian Institute of Astrophysics, Koramangala, Bangalore, India \\}

\author{Selvakumar G}
\affiliation{Indian Institute of Astrophysics, Koramangala, Bangalore, India \\}

\author{Maheswer G}
\affiliation{Indian Institute of Astrophysics, Koramangala, Bangalore, India \\}

\author[0000-0002-4999-2990]{R. Arun}
\affiliation{Department of Physics and Electronics, CHRIST (Deemed to be University), Hosur Main Road, Bangalore, India\\}
\affiliation{Indian Institute of Astrophysics, Koramangala, Bangalore, India \\}

\author{Hema Anilkumar}
\affiliation{Department of Physics and Electronics, CHRIST (Deemed to be University), Hosur Main Road, Bangalore, India\\}

\author[0000-0001-8873-1171]{Gourav Banerjee}
\affiliation{Department of Physics and Electronics, CHRIST (Deemed to be University), Hosur Main Road, Bangalore, India\\}

\author {Pramod Kumar S}
\affiliation{Indian Institute of Astrophysics, Koramangala, Bangalore, India \\}

\author[0000-0002-7666-1062]{Sreeja S Kartha}
\affiliation{Department of Physics and Electronics, CHRIST (Deemed to be University), Hosur Main Road, Bangalore, India\\}

\author{KT Paul}
\affiliation{Department of Physics and Electronics, CHRIST (Deemed to be University), Hosur Main Road, Bangalore, India\\}

\author{C. Velu}
\affiliation{Indian Institute of Astrophysics, Koramangala, Bangalore, India \\}






\begin{abstract}
We present a follow-up study on the recent detection of two X-ray flaring events by MAXI/GSC observations in soft and hard X-rays from MAXI J0709-159 in the direction of HD 54786 (LY CMa), on 2022 January 25. The X-ray luminosity during the flare was around $10^{37}$ erg/s (\textit{MAXI}), which got reduced to $10^{32}$ erg/s (\textit{NuSTAR}) after the flare. We took low-resolution spectra of HD 54786 from HCT and VBT facilities in India, on 2022 February 1 and 2. In addition to H$\alpha$ emission, we found emission lines of He{\sc i} in the optical spectrum of this star. By comparing our spectrum of the object with those from literature we found that He{\sc i} lines show variability. Using photometric study we estimate that the star is having effective temperature of 20000 K. Although HD 54786 is reported as a supergiant in previous studies, our analysis favours it to be evolving off the main sequence in the Color-Magnitude Diagram. We could not detect any infrared excess, ruling out the possibility of IR emission from a dusty circumstellar disc. Our present study suggests that HD 54786 is a Be/X-ray binary system with a compact object companion, possibly a neutron star.
\end{abstract}





\section{Introduction} \label{sec:intro}

Be/X-ray binaries (BeXRBs) form a major subclass of high-mass X-ray binaries which consist of a Be star and a compact object \citep[e.g.][]{2011Reig,  1982Rappaport}. The compact object accretes matter from the decretion disc of the Be star. The possible compact objects are neutron star (NS) \citep{2001OkazakiAndNegueruela}, white-dwarf (WD) \citep{2021Kennea} and black hole (BH) \citep{2014Casares}, yet NSs are most frequently observed companion than other kinds \citep{2009Belczynski}. BeXRBs can be persistent or transient in nature. Whereas in the latter case it exhibits outburst events, which are classified as Type I, where X-ray luminosity $L_x$$\leq$10\textsuperscript{37}erg s\textsuperscript{-1} and occur regularly, separated by the orbital period, and Type II, where L$_x$$\geq$10\textsuperscript{37}erg s\textsuperscript{-1}, this outbursts are less frequent and are not modulated on the orbital period \citep{1986Stella}. The transient X-ray behavior is caused by the interaction of the compact object with the material in the circumstellar disc \citep{2017Monageng, 2001OkazakiAndNegueruela}. A Be star is a special class of massive B-type main sequence star, surrounded by a geometrically thin, equatorial, gaseous, decretion disc \citep{2007Meilland}. Be stars are primarily identified by means of H$\alpha$ emission feature formed in the ionized decretion disc of Be stars, by means of the recombination process \cite[e.g.][]{2021Banerjee, 2012Mathewxper}.

Recently, on 2022 January 25 at 10:42:28 UT, the Monitor of All-sky X-ray Image (MAXI) Gas slit camera (GSC) \citep{2016Negoro} nova alert system triggered a bright uncatalogued X-ray transient source with an X-ray flux of 266 $\pm$ 32 mCrab over the scan. The source of flare is situated at an RA of 107.280 deg and Dec of -15.923 deg. The intensity of X-ray flux was comparatively low before (09:09 UT) and after (12:15 UT) the event \citep{2022ATelserino}. The source was named as MAXI J0709-159. Following that, at 13:44 UT (on the same day), Neutron Star Interior Composition Explorer Mission \citep[NICER;][]{nicer2016SPIE.9905E..1HG} reported X-ray emission from the position of RA = 107.392 deg and Dec = -16.108 deg, with an X-ray flux of 10 mCrab in the energy band of 0.2$-$12 keV \citep{2022ATelIwakiri}.

\cite{2022ATelKobayashi} found that the position of the Be star HD 54786 matches with that of the region from where X-ray emission is detected by NICER. Based on the Gas Slit Camera (GSC) energy spectrum study, they suggested HD 54786 to be a companion of MAXI J0709-159, and that the system might be a supergiant Fast X-ray Transient (SFXT) or a Cir X-1 type object. This was further corroborated from the study of \cite{2022ATelNegoro}, who analyzed the X-ray emission using the Nuclear Spectroscopic Telescope Array \citep[NuSTAR;][]{NEUSTAR2013ApJ...770..103H} observations from 00:21 to 11:21 UT on 2022 January 29.

The first spectroscopic study of HD 54786 is by \cite{1949Merrill} who identified hydrogen emission lines in the spectrum. \cite{1955Morgan} classified this star as a blue-supergiant with luminosity class I. Subsequently, it was listed in the Be star catalog of \cite{1982Jaschek}. Later, \cite{2015Chojnowski} also identified this star as a Be star. \cite{1988Houk} estimated its spectral type to be B1/B2.

This letter follows up on the recent X-ray flaring event of HD 54786, combining optical spectroscopy and photometry over various epochs, to better understand the nature of the prospective BeXRB system.

\begin{figure*}
\centering
   \includegraphics[width=\linewidth,height=\textheight,keepaspectratio]{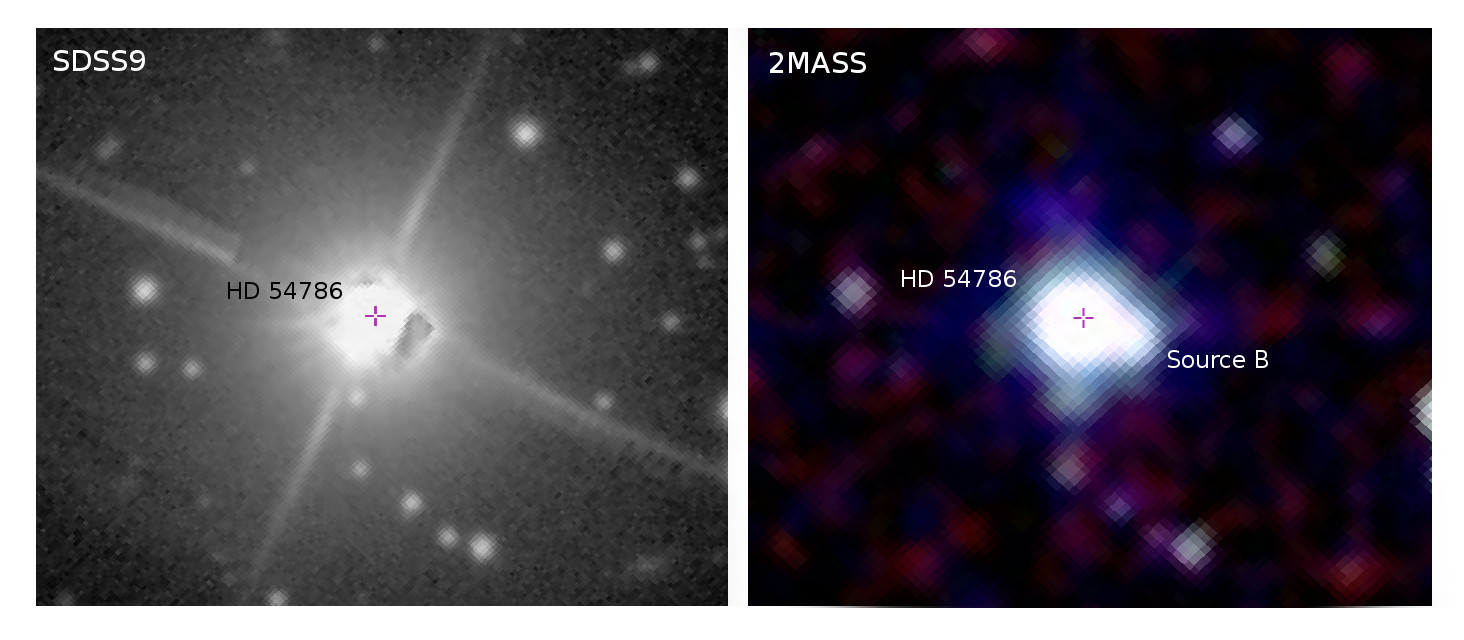}
    \caption{Figure shows images of the region around HD 54786 (FoV 1.3') from SDSS9 survey (left) in grey scale and 2MASS survey (right) in J, H and K$_S$ color composite, respectively. The source B is distinguishable in the 2MASS image but not in the SDSS9 image.}
    \label{2massrep}
\end{figure*}

\section{Observations}
\subsection{Optical spectroscopy}

We collected two low resolution spectra of HD 54786, from HFOSC instrument mounted on the 2.01-m Himalayan Chandra Telescope (HCT)\footnote{http://www.iiap.res.in/iao/hfosc.html} located at Hanle, Ladakh, India. We observed the star on two consecutive dates, 2022 February 01 \& 02. The spectral coverage is from 5500$-$8800 \AA, where both the red region spectra are obtained with Grism 8 (5800$-$8300 \AA) and 167 $\mu$m slit, providing an effective resolution of 7 \AA~at H$\alpha$.

To validate the spectra obtained from HCT, we acquired two more simultaneous low resolution spectra at the same epochs from the OMR instrument, mounted on the 2.34-m reflecting telescope situated at the Vainu Bappu Observatory (VBO), Kavalur, Tamil Nadu, India. In VBT the photographic sensor employed is an Andor CMOS high speed read out sensor, comprising of 1024 $\times$ 256 pixels having pixel size of 26 $\mu$m. The spectra are obtained in the wavelength region of 5500$-$8500 \AA, at settings particularly centered at H$\alpha$ line with a resolving power of 1000.

For both observations, dome flats taken with halogen lamps were utilized for flat-fielding the images. Bias subtraction, flat-fielding and spectrum extraction were performed using IRAF routines. The wavelength calibrated spectra are continuum fitted using IRAF tasks. Additionally, the resulting spectra from VBT are further normalized using python routine.

\subsection{\textit{Gaia} astrometric data}
The astrometric data of HD 54786 is taken from the latest data release of Gaia (\textit{Gaia} EDR3; \citealp{GaiaCollaboration2021A&A...649A...1G}). The geometrical distance estimates by \cite{2021Bailerjonesdr3} is used in this study. The CMD analysis using the \textit{Gaia} EDR3 data has given our work a greater context (section \ref{CMD}).

\subsection{X-ray}

HD~54786 has been continuously monitored with \textit{MAXI} for years. As the recent \textit{MAXI} archival data were not available, we used the publicly available processed GSC (2$-$30~keV) light curves (version 7Lrkn 4h), which is frequently updated every 4 hours. \textit{NuSTAR} also observed the source during the quiescent period for 18~ks on 2022 January 29. The two co-aligned focal plane modules, FPMA \& FPMB, detected hard X-ray photons (3$-$79~keV) from the source. We generated the spectra and light curves from FPMA and FPMB observations using standard pipelines in the \textit{NuSTAR} Data Analysis Software (NUSTARDAS V2.0.0; CALDB Version 20220131). The source spectra were extracted for circular regions of radii 32.32\arcsec while the background spectra were generated for annular regions (r$_{in}$=32.32\arcsec, r$_{out}$=60.00\arcsec) centered on the source.

\section{Results}

\subsection{\textit{Gaia} EDR3 analysis}
The distance of HD 54786 was reported to be different in previous studies. So, we used the \textit{Gaia} EDR3 distance estimate of $2955^{+244}_{-175}$ pc \citep{2021Bailerjonesdr3} for further analysis.

From the \textit{Gaia} database and the field image of 2MASS (Fig. \ref{2massrep}), we note that there is another star (source B) at a separation of 6.5" in the south-west direction of HD 54786. This is not prominent in the optical image (e.g. SDSS9) but detectable in 2MASS J,H and K$_S$ composite image. The astrometric parameters such as parallax and proper motions of two stars must be similar and within uncertainty limits for being considered to be in a binary system \citep{Arun2021MNRAS.501.1243A}. We found that the distances of the two sources are matching within uncertainties. However, their proper motion values are considerably different. Hence, for the present study we will not consider them as associated and further discussion will purely be on HD 54786.

\subsection{\textit{Gaia} CMD analysis}
\label{CMD}

\begin{figure*}
\centering
   \includegraphics[width=\linewidth,height=\textheight,keepaspectratio]{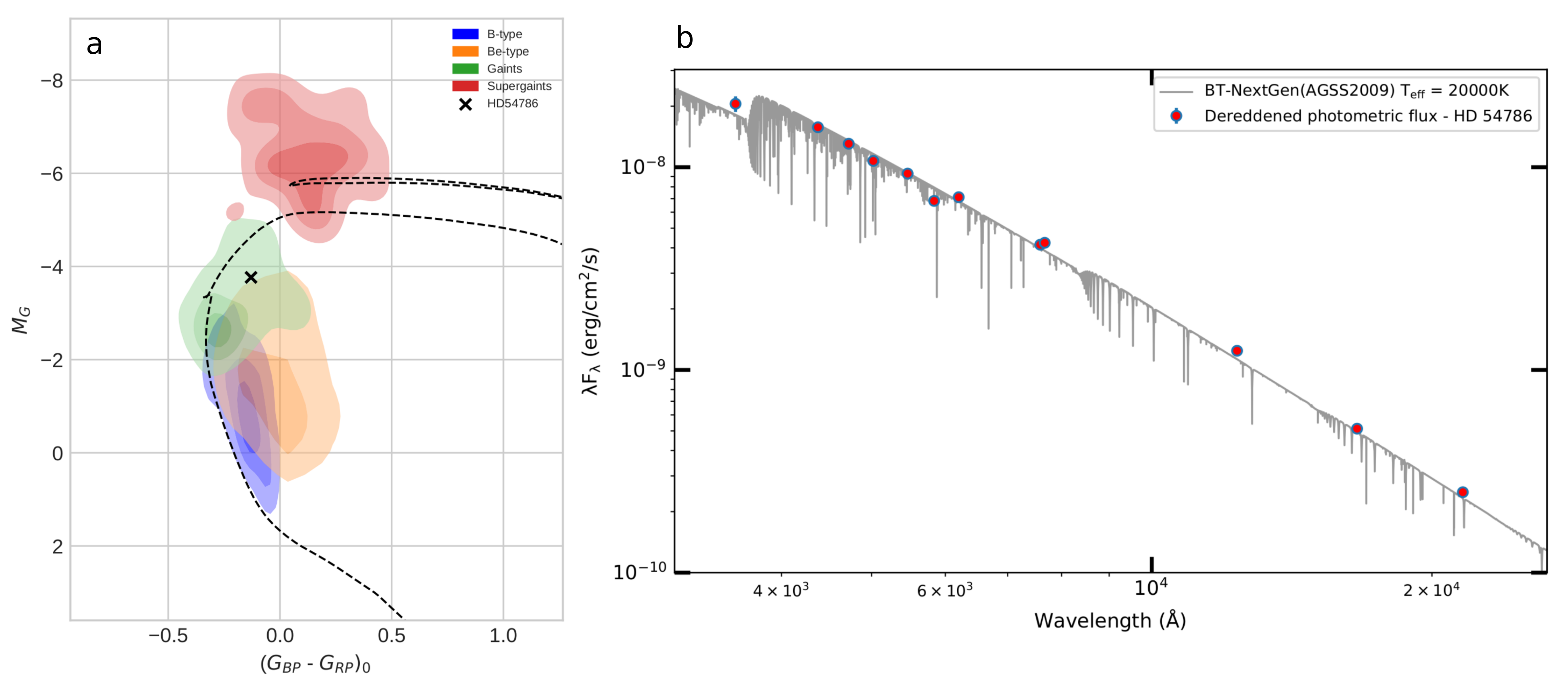}
    \caption{Panel (a): The \textit{Gaia} CMD of HD 54786 having absolute \textit{Gaia} G and color corrected (BP-RP) magnitudes available from \protect\cite{GaiaCollaboration2021A&A...649A...1G}. The probability distribution (Gaussian fitted at three contour levels) of the B, Be stars, Giants and supergiants are shown in blue, orange, green and red shaded colors, respectively. The black dashed line in the plot represents the isochrone of 60 Myr with V/Vcrit = 0.4 and [Fe/H] = 0 (The top black dashed line is the blue loop part of the same isochrone). Panel (b): In the SED, the flux values of the star HD 54786 is fitted with the theoretical BT-NextGEN(ASGG2009) model at T\textsubscript{eff} = 20000 K and log(g) = 3, using the chi-squared minimization method.}
    \label{gaiacmd}
\end{figure*}

HD 54786 is reported as a supergiant by \cite{1988Houk}, belonging to the luminosity class Ib. We constructed the \textit{Gaia} color-magnitude diagram (CMD) for this star to identify its location. Considering its \textit{Gaia} EDR3 distance and A\textsubscript{V} value of 0.93 (as estimated from the Green's map; \citealp{2019Green}), we plotted the corresponding $M_G$ versus $G_{(BP-RP)0}$ CMD for the star. Additionally, we over-plotted the probability distribution (Gaussian fitted at three contour levels) of previously studied B-type \citep{2010Huang}, Be stars \citep[and references therein]{2021Bhattacharyya}, Giant stars \citep{Hohle2010AN....331..349H} and supergiants \citep{2021Georgy} to better understand the location of HD 54786. From \cite{2013pecautmamajek}, the ZAMS (zero-age main sequence) line does not extend beyond B9 spectral type in $M_G$ and $G_{(BP-RP)0}$ values. Hence, we utilised the closely matching 60 Myr isochrone track from MESA (Modules for Experiments in Stellar Astrophysics) isochrones and evolutionary tracks (MIST) \citep{2016choi, 2016dotter}, which is over-plotted in the CMD with V/V$_{crit}$ = 0.4, since that is the only model available in the MIST database for a rotating system. Moreover, we adopted the metallicity value of [Fe/H] = 0,  corresponding to solar metallicity $Z_\odot$ = 0.0142 \citep{2009Asplund} for the tracks. The CMD showing the location of HD 54786, with the representative locations of main sequence B-type stars, Be stars, giants and supergiants is presented in Fig. \ref{gaiacmd}a.

Our analysis shows that the location of HD 54786 is near to the top of the distribution of B and Be stars. It is also noticed that the star resides inside the distribution of giant stars but below that of supergiants. This indicates that HD 54786 might be an evolved star with respect to B/Be types. However, its location also conveys that it is not a supergiant star.

\subsection{Spectral energy distribution}
As a next step, we fitted the spectral energy distribution (SED) for HD 54786 to re-estimate its spectral type. For generating the SED, we used Johnson U magnitude, B, V, g', r', and i' magnitudes from APASS \citep{2015APASS}; G, G\textsubscript{RP} and G\textsubscript{BP} magnitdes from \textit{Gaia} EDR3 and the infrared J, H and K\textsubscript{s} magnitudes from 2MASS. The SED is generated with the python routine used in \cite{2021Arunb}. Fig. \ref{gaiacmd}b presents the SED plot generated for HD 54786. From a grid of BT-NextGen (AGSS2009) \citep{1999Hauschildt} theoretical stellar atmospheres at solar metallicity and log(g) = 3$-$4 (since it seems to be evolved from the main sequence in the CMD), corresponding to various temperatures, we found that the best fit (by means of chi-squared minimization) corresponds to an effective temperature ($T_{eff}$) of 20000 $\pm$ 500 K with log(g) of 3.

The obtained $T_{eff}$ matches with a spectral type of B2 \citep{2013pecautmamajek}, if it is a main sequence (MS) star. However, the stellar radius estimated from the SED is R\textsubscript{*}/R\textsubscript{$\odot$} = 11.8. This value is higher than that of a MS star (5.7 R\textsubscript{$\odot$}) with a similar $T_{eff}$ \citep{2013pecautmamajek}. \cite{Allen2000asqu.book.....C} reported that a supergiant with the similar spectral type will have a radius of $\sim$ 19 R\textsubscript{$\odot$}. Our estimated log(g) and stellar radius values suggest that the star is not on the MS phase, but is in the evolved region (off the MS), which is in agreement with our CMD analysis. Hence, we will not resort to identifying any specific spectral type for this star in the present work. Instead, we agree that the spectral type is similar to the previous estimates of B1/B2 \citep{1988Houk}, and we will put forth the $T_{eff}$ and log(g) value as the derivatives from this study.

Furthermore, we do not notice any significant IR excess in the SED, suggesting the absence of dust emission from the circumstellar disc. If the disc contains a dust component, it should appear as flux excess in the 2MASS colors. The lack of dusty disc eliminates the possibility of the extended accretion disc. However, based on the H$\alpha$ emission, a decretion disc surrounding the Be star is possible, with the excess caused by free-free emission or thermal bremsstrahlung, which is more apparent in the mid-IR range. Although mid-IR data is available in ALLWISE survey \citep{2014cutri}, we did not consider it since there is a nearby star that will contaminate the WISE magnitudes.

\subsection{Prominent spectral features identified in HD 54786}
\begin{figure*}
\centering
   \includegraphics[height=105mm, width=142mm]{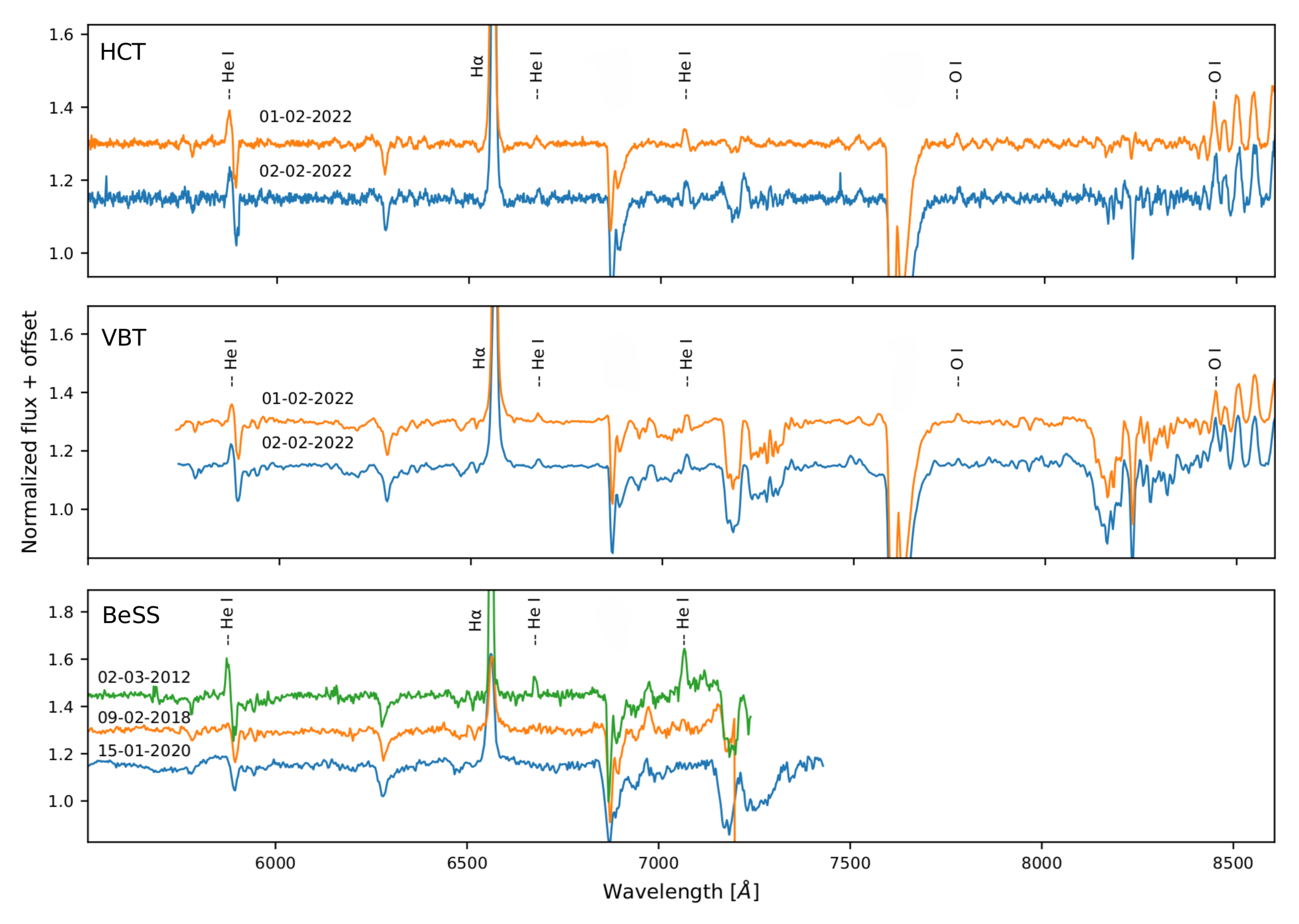}
\caption{Representative spectra of HD 54786 showing different spectral features in the wavelength range of 5500 - 8500 {\AA}. The spectra from HCT, VBT and BeSS spectrum are listed in top, middle and lower panel.}
\label{fig5rep}
\end{figure*}

We then identified all major spectral features observed in the spectra of HD 54786 taken using VBT and HCT. We found that the H$\alpha$ line is visible in single-peak emission in all spectra, for each of the epochs. The average signal-to-noise ratio (SNR) of every spectrum taken is greater than 120. In case of HCT, we detected that H$\alpha$ equivalent width (EW) was -17.6 \AA~on 2022 February 01, whereas it was -16.9 \AA~on 2022 February 02. The H$\alpha$ EW from HCT spectra decreased by a value of 0.7 \AA~within one day. Fig. \ref{fig5variation} (top panel) represents the H$\alpha$ EW variation with time. Apart from H$\alpha$, emission lines of He{\sc i}: 5876, 6678, 7065, O{\sc i}: 7772, 8446 \AA~and P13 - P17 of the Paschen series are also noticed on both the dates. In addition, we also observed telluric O$_2$ absorption bands of 6867 and 7593 \AA~ \citep{2015Smette} on both the cases. Fig. \ref{fig5rep} shows the representative spectra of HD 54786 obtained using HCT and VBT.

Interestingly, H$\alpha$ EW of HD 54786 was found to be -23 \AA~when observed on 2022 January 28 by \cite{2022ATelNesci}, which is taken 3 days after the X-ray flaring. It is intriguing to notice that the H$\alpha$ EW was higher on the nights closer to the flaring event and is decreasing to -16.9 \AA ~on 2022 February 02. However, any major change of EW is not noticed for any other emission line feature other than H$\alpha$, during the period of our observations.

\subsubsection{Spectral features observed in data from the BeSS database}
We queried in the BeSS database \citep{2011Neiner} to check whether any observation of this star was done by amateur astronomers. We found a total of three spectra for HD 54786 taken by amateur astronomers Dejean Pastor (spectrum taken on 2020 January 15), Cazzato (2018 February 09) and Buil (2012 March 02), in the BeSS database. The spectra are represented in Fig. \ref{fig5rep}.

It was found that H$\alpha$ EW was the maximum (around -28.3 \AA) on 2012 March 02. This value decreased to -2.7 \AA~on 2018 February 09. Interestingly, H$\alpha$ EW again increased to -12.1 \AA~on 2020 January 15. H$\beta$ is also seen in emission (EW = -3.1 \AA) on 2012 March 02, whereas it shows weak absorption on the other two dates. Moreover, He{\sc i} 5876, 6678 and 7065 \AA~lines are noticed in emission only on 2012 March 02, whereas in the other two occasions (2018 February, 09 and 2020 January, 15, respectively) we are not seeing these three lines in emission.

\subsubsection{Variability in He{\sc i} emission lines}
Comparing the spectra obtained by us using the HCT facility and those retrieved from the BeSS database, we found a considerable variability exhibited by He{\sc i} 5876, 6678 and 7065 \AA~emission lines. All these lines were visible in emission on 2012 March 02 when amateur astronomer Buil observed HD 54786 (C11 LISA ATIK314L+). On the contrary, all were present in absorption in the BeSS spectra on 2018 February 09 and 2020 January 15. However, we found all of them in emission when observed with HCT on 2022 February 01 and 02.

It is known that He{\sc i} emission lines are found rarely in Be stars \citep{2021Banerjee} and are produced in high temperature regions with T $\sim$ 15,000 K \citep{2011Kwan}. \cite{1975Bahng} proposed that these lines may either be a temporary phenomenon or non-LTE effects are responsible for their selective excitation. Since we find that He{\sc i} lines are showing intensity changes over a short period of time, it may be a temporary phenomenon, as proposed by \cite{1975Bahng}. \cite{1994Apparao&Tarafdar} proposed that in Be stars, He{\sc i} emission lines can be produced by the influence of a compact binary companion which is accreting matter from the disc of the primary (here the Be star) star.

The occasional appearance and disappearance of He{\sc i} emission lines in the series of spectra used in our study indicate that the disc temperature of HD 54786 becomes higher during the flaring event and has a recurring nature. However, appearance of these lines in emission on 2022 February 01 and 02 hints towards some sudden increase in disc temperature. This can happen if some flare activity takes place from a nearby companion star, which is what has been observed recently in case of HD 54786. Hence, our study suggests the possibility that X-ray flaring probably resulted in the formation of He{\sc i} emission lines, which may disappear as time progresses. We did not see such a correlation earlier since X-ray data was not available then. We are doing follow-up observations for HD 54786 to check whether any X-ray flaring event leads to the formation of He{\sc i} emission lines or not.

\subsection{X-ray spectral analysis}
\begin{figure}[h]
\centering
   \includegraphics[width=\linewidth,height=\textheight,keepaspectratio]{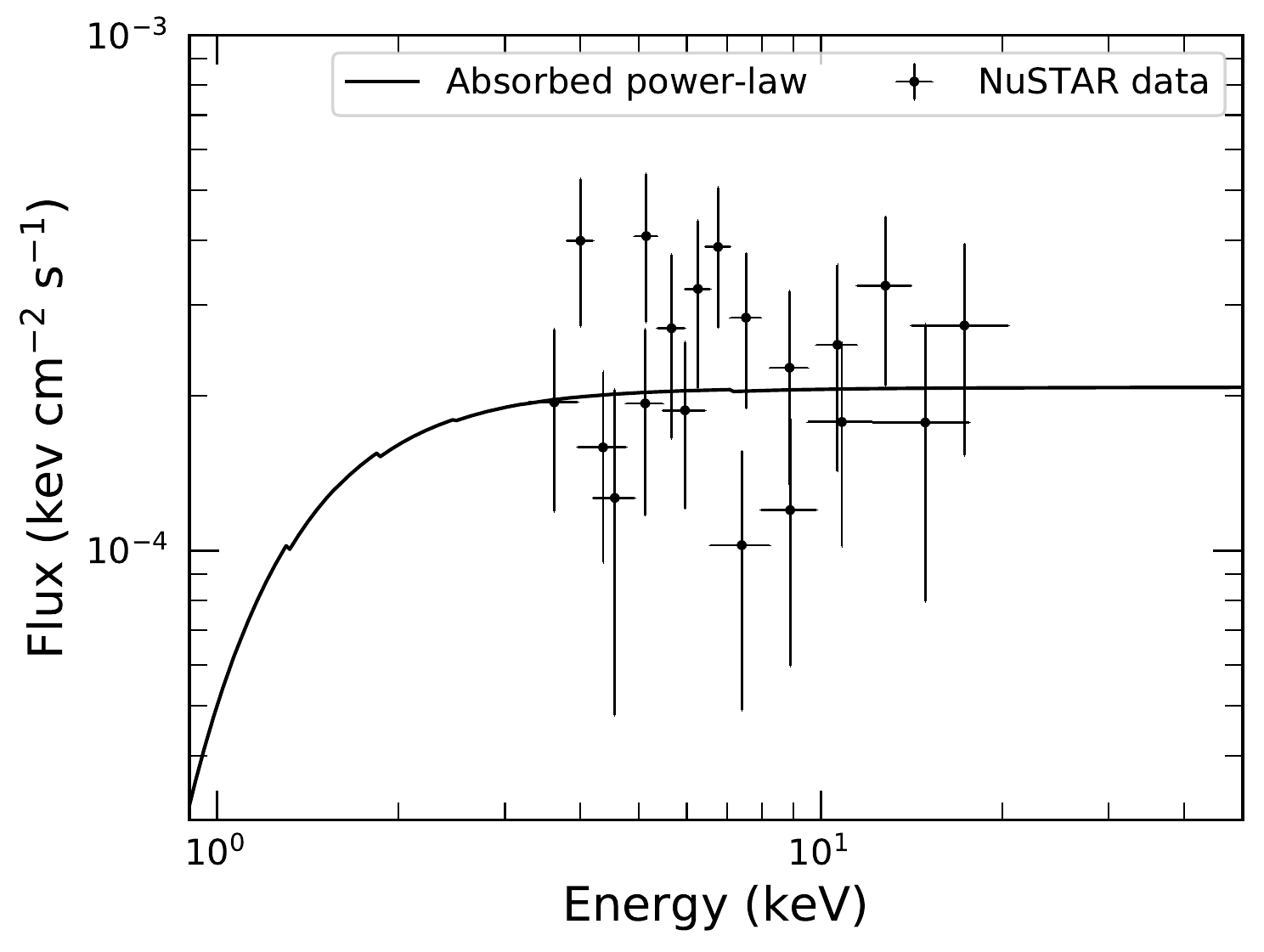}
\caption{
\textit{NuSTAR} (FPMA \& FPMB) spectra modelled with $constant \times TBabs \times powerlaw$ in the 3--25~keV band. The solid line shows the power-law continuum corrected for the Galactic absorption.
}
\label{fig-xray}
\end{figure}

We performed the X-ray spectral analysis of the NuSTAR observation of the source for the quiescent period using the X-ray spectral fitting package \emph{Xspec} \citep[version 12.12.0;][]{1996Arnaud}. The spectra were grouped to have a minimum of 15 counts per bin. Above 25~keV, the spectra were dominated by the background. As a result, we looked at the spectra in the 3$-$25~keV energy range.
We used an empirical power-law model modified by galactic absorption to fit the spectra. The galactic absorption was modelled with the neutral hydrogen column density $N_H$ fixed at ${\rm 5.6\times10^{21}~cm^{-2}}$, obtained using the HI4PI survey \citep{2016HIP4PIBkhti}. We did not consider the intrinsic absorption as it was not considerably significant. 
To account for the cross-normalisation disparities between the FPMA and FPMB data, a constant factor ($0.94^{+0.37}_{-0.28}$) was applied. A photon index (slope of the hard X-ray power-law) of $\Gamma=2.00^{+0.47}_{-0.43}$ was obtained from the model fit with a fit-statistic of $\chi^2$(degrees of freedom)=18.26(16). To measure the unabsorbed flux we added another component \emph{cflux} to the power law model. We measured the unabsorbed X-ray flux in the energy range of 3$-$25~keV to be ${\rm 7.03^{+1.81}_{-1.68} \times 10^{-13}erg~cm^{-2}s^{-1}}$. The best-fit model and unfolded spectra are shown in Fig~\ref{fig-xray}.

The X-ray (2$-$10~keV) flux values of the source for the flaring period are obtained from \cite{2022ATelKobayashi}. We then calculated the luminosity of the two detected peaks during the flaring, which was reported by MAXI/GSC at 10:42 and 13:48 UT on 2022 January 25. The luminosity for the first and second flares are estimated to be $6.42^{+2.82}_{-1.67} \times 10^{36}$ erg s\textsuperscript{-1} and $1.67^{+4.29}_{-1.05} \times 10^{37}$ erg s\textsuperscript{-1}, respectively. Moreover, we calculated the unabsorbed 2--10~keV flux from the \textit{NuSTAR} data from 00:21 to 11:21 UT on 2022 January 29, about four days after the initial MAXI detection. The spectral fit is found to be ${\rm 5.33^{+1.47}_{-1.32} \times 10^{-13}erg~cm^{-2}s^{-1}}$ and the X-ray luminosity, for a distance of 2955 pc, to be $5.57^{+1.54}_{-1.38} \times 10^{32}$ erg s\textsuperscript{-1}.

\subsection{Is HD 54786 a $\gamma$ Cas variable or BeXRB?}
Based on irregular photometric variability, HD 54786 is reported as a $\gamma$ Cas variable  \citep{2012Alfonzo}. However, it is difficult to differentiate $\gamma$ Cas variables and BeXRBs based on optical variability alone. They can differ significantly depending on their X-ray characteristics \citep{2011Reig}. In this section we evaluate whether HD 54786 is a $\gamma$ Cas variable or a BeXRB binary system.

Typically, $\gamma$ Cas variables (mostly a Be + white dwarf system) show harder X-ray spectra, best fit with a thin plasma emission model \citep{2016Smith}. On the contrary, a power law model is usually observed in high mass X-ray binaries (HMXBs), where BeXRB is a subclass. Moreover, $\gamma$ Cas variables do not exhibit large X-ray outbursts \citep{2011Reig}. The hard X-ray emission from BeXRBs are usually non-thermal in nature, which is not the case in $\gamma$ Cas variables \citep{1998PASJKubo50..417K}.

\begin{figure*}
\centering
   \includegraphics[width=\linewidth,height=\textheight,keepaspectratio]{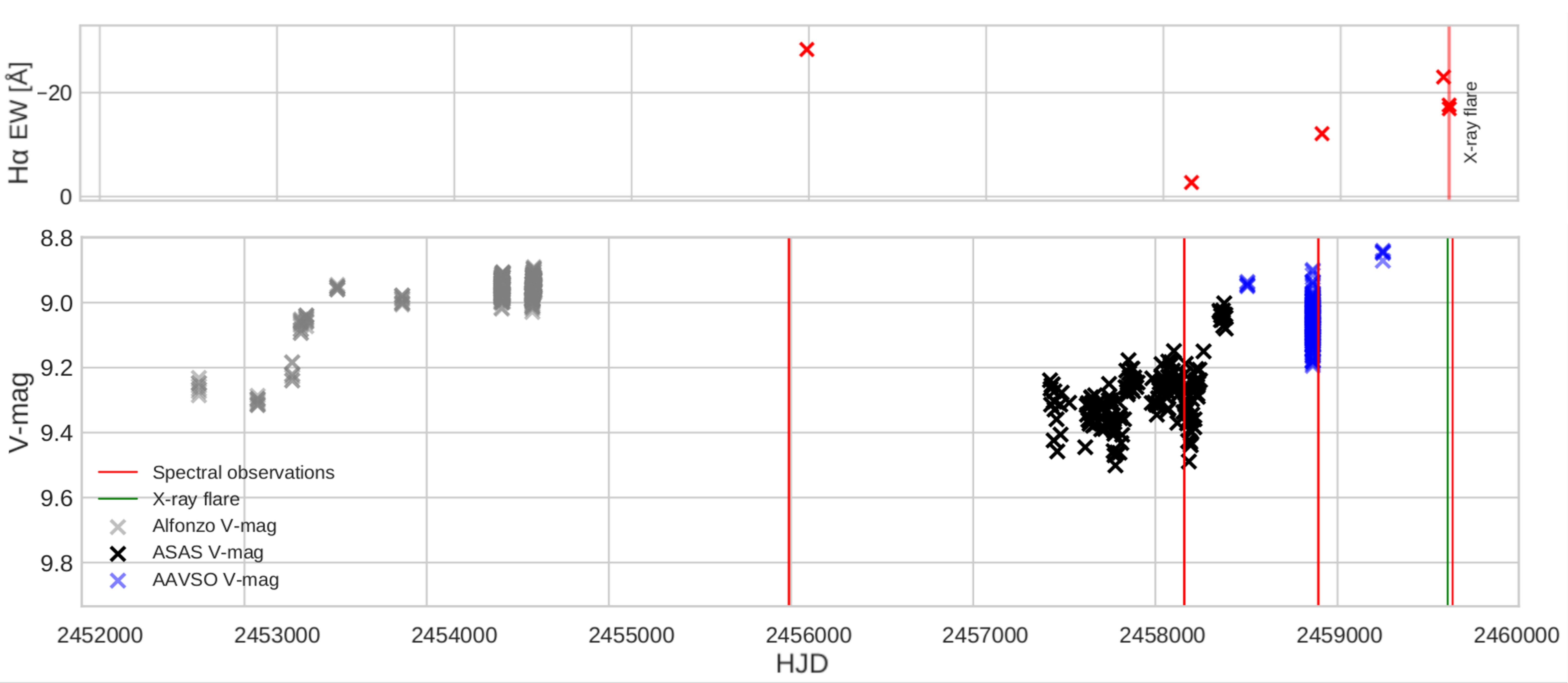}
\caption{Variation of $H\alpha$ EW and V-mag shown by HD 54786. In the top panel, the red cross marks represent the H$\alpha$ EW. In the bottom panel, the grey, black and blue cross marks show V-mag values obtained from \cite{2012Alfonzo}, ASAS and AAVSO, respectively. The vertical red lines and green line represent the time of spectral observations and the X-ray flaring event, respectively. It is noticed that its V-mag steadily increases over the span of observations. The epoch of BeSS spectra from 2018 February (HJD 2458150) and the photometry measurements from ASAS are seen to overlap in the figure. During that time, the $H\alpha$ EW was significantly lower (-2.7 \AA), with an average V-mag of 9.2$-$9.3. Again from 2018 May$-$2018 September (HJD 2458239 to 2458362), V-mag values were between 9.0 and 9.1. Then, a considerable rise in $H\alpha$ EW (to -12.1 \AA) in the BeSS spectra of 2020 January (HJD 2458849) was detected, whereas AAVSO V-mag value was found to range between 8.9 and 9.2. Subsequently, the V-mag changed to 8.85 during AAVSO observations in 2021 February (HJD 2459246). However, a considerable variability of $H\alpha$ EW is seen, i.e., from -23.2 to -16.9 \AA, if we compare our observations with that of \cite{2022ATelNesci}.}
\label{fig5variation}
\end{figure*}

For HD 54786, the X-ray luminosity of the source was found to be of the order 10\textsuperscript{37} erg s\textsuperscript{-1}, when observed by MAXI/GSC on two different times on the same date, i.e., on 2022 January 25. The luminosity then dropped to 10\textsuperscript{32} erg s\textsuperscript{-1} on 2022 January 29, when observed by NuSTAR. The X-ray flare of such a scale suggests that HD 54786 belongs to the class of BeXRB, where the companion is a compact object, possibly a neutron star \citep[e.g.][]{2011Reig, 2001OkazakiAndNegueruela}. The X-ray luminosity as estimated by us implies that the outburst is of Type I \citep{1986Stella}. However, further study using time period analysis and periodic pulsations will be necessary to confirm this.

\section{Discussion}
Our study suggests that HD 54786 might be an evolved Be star rather than a main sequence star or a supergiant. Using SED analysis we estimated its T$_{eff}$, log(g) and radius (R\textsubscript{*}/R\textsubscript{$\odot$}) values to be 20000 K, 3 and 11.8, respectively. We found a considerable decrease in H$\alpha$ EW from -23 \AA~on 2022 January 28 (noticed by \cite{2022ATelNesci} 3 days after the X-ray flare) to -16.9 \AA~on 2022 February 02. It is interesting to notice that the H$\alpha$ EW was higher on the nights closer to the flaring event and decreased on later dates. The variability we observed in He{\sc i} 5876, 6678 and 7065 \AA~emission lines indicate that the disc temperature of HD 54786 becomes higher during the flaring event and has a recurring nature. 

Our results further suggest the possibility that X-ray flaring probably resulted in the formation of He{\sc i} emission lines, which may disappear as time progresses. Moreover, the X-ray analysis points out that the outburst is of Type I and HD 54786 belongs to the class of BeXRB consisting a compact object companion.

Furthermore, we studied the epoch-wise variation of V-mag and $H\alpha$ EW seen in HD 54786 in the context of the X-ray flare event. This is done to check whether the variation of V-mag and $H\alpha$ EW is co-related with the X-ray flaring event or not. For that, we searched the \textit{MAXI} on-demand webpage for the epochs where a variation in H$\alpha$ was observed (2012, 2018 and 2020), and did not find any significant X-ray detection during these periods. Also, we compiled the V-magnitude values from \cite{2012Alfonzo} (between April 2003 and May 2008), ASAS-SN Catalog of Variable Stars III \citep{2014Shapee, 2019Jayasinghe} (2016 February to 2018 September), and from the AAVSO database, for the period from 2019 January to 2021 February. Fig. \ref{fig5variation} presents the variation of $H\alpha$ EW and V-mag shown by HD 54786. Interestingly, our study shows that there was a gradual increase in V-mag which took place till 2021 February, the last date of observation available before the X-ray flaring event. Further observational data, particularly simultaneous observations are required to understand whether the V-mag and $H\alpha$ emission are associated with the X-ray outburst or not. Also, it will help to assess the periodicity of the HD 53786 BeXRB system.

\section{Acknowledgements}
We would like to thank the Center for research, CHRIST (Deemed to be University), Bangalore, India for providing the necessary support. Moreover, we convey our heartiest gratitude to the staff of the Indian Astronomical Observatory (IAO), Ladakh and Vainu Bappu Observatory, Kavalur for taking the immediate observations using the HCT and VBT, respectively. R.A is grateful to the Centre for Research, CHRIST (Deemed to be University), Bangalore for the research grant extended to carry out the present project (MRP DSC-1932). B.M acknowledges the support of the Science \& Engineering Research Board (SERB), a statutory body of the Department of Science \& Technology (DST), Government of India, for funding our research under grant number CRG/2019/005380. This study has used the \textit{Gaia} EDR3 data to adopt the corresponding distances for program stars. Hence, we express our gratitude to the Gaia collaboration. We also thank the SIMBAD database and the online VizieR library service for helping us in the relevant literature survey. Furthermore, this work has made use of the BeSS database, operated at LEISA, Observatory de Meuden, France (http://basebe.obspm.fr). Hence, we thank the BeSS database too.

\appendix
\restartappendixnumbering
 \begin{table*}[h]
    \caption{Observation log of HD 54786 from HCT and VBT}\label{tab1features}

    \begin{tabular}{lllllll}
    \hline
    Star Name & Date and time(UT) of observation & HJD   & RA      & DEC       & Exposure Time(s)                 \\ 
    & (yyyy-mm-dd, hh:mm) &&&&& \\ \hline
    HCT Observation \\ \hline
    HD 54786  & 2022-02-01, 16:22  & 2459611 & 07 09 36.97 & -16 05 46.1 & 240 \\
    & 2022-02-02, 13:39  & 2459612 & - & - & 180 \\
    \\ 
    \hline
    VBT Observation \\ \hline 
    HD 54786  & 2022-02-01, 16:00  & 2459611 & 07 10 46.97 & -16 07 46.1 &  1800\\
    & 2022-02-02, 15:08  & 2459612 & - & - & 2400 \\ \\\hline

    \hline
\end{tabular}
\end{table*}


\bibliography{sample631}{}
\bibliographystyle{aasjournal}



\end{document}